\documentclass[nofootinbib,10pt,aps,pra,twocolumn]{revtex4}   

\usepackage[dvips]{epsfig}
\usepackage{amsmath}    
\usepackage{amsfonts}  
\usepackage{amssymb}
\usepackage{graphicx}   
\usepackage{mathrsfs}

\usepackage[ps2pdf,colorlinks]{hyperref}

\AtBeginDocument{%
    \hypersetup{
      urlcolor     = blue, 
      linkcolor    = black, 
      citecolor   = blue, 
      pdfborder={0 0 0},
      urlbordercolor={1 1 1},
      citebordercolor={1 1 1}
    }
    }

\begin{document}

\title{The quantum optical description of a double Mach-Zehnder interferometer}
\author{Stefan~Ataman}
\affiliation{ECE Paris, 37 quai de Grenelle, 75015 Paris, France}
 \email{ataman@ece.fr}

\begin{abstract}
In this paper we describe within the formalism of Quantum Optics
(QO) a double Mach-Zehnder interferometer (MZI). For single photon
Fock states this experimental setup is shown to exhibit a
counter-intuitive behavior: for certain values of the path length
difference of the first MZI, the singles photon-count statistics
at the output detectors does not change, whatever the difference
in path length for the second MZI. For simultaneously impinging
light quanta, we show that this setup is able to show the same HOM
antibunching effect previously obtained with a beam splitter.
However, by adding substantial delays in each MZI, we can obtain
the same effect even if the ``photon wave packets'' do not overlap
at the second beam splitter.
\end{abstract}


 \maketitle

\section{Introduction}

The beam splitter (BS) is one of the most widely used devices in
Quantum Optics. This notoriety is partially due to the fact that a
beam splitter can transform a non-entangled state into an
entangled one \cite{Lou03,Man95,Ger04,Gry10} (and vice-versa),
differentiate between a coherent and a Fock state
\cite{Gra86,Asp91} and reveal non-classical features of light
\cite{Kim77,Gho87,HOM87}. Applying a coherent state
\cite{Gla63,Sud63,Gla63b} at one input and a single-quantum Fock
state at the other one of a BS allows measuring quantum states of
light using the homodyne detector \cite{Yue78,Yue83,Leo95}.

In the classical description of a lossless beam splitter, energy
conservation imposes the relation between input and output
electric fields \cite{Lou03,Gry10}. In the quantum optical
description, fields are replaced by operators \cite{Tan97}. A more
general framework has been developed, where beam-splitters are
described in SU(2) symmetries \cite{Yur86,Cam89}. However, authors
prefer a subset of this general model, having symmetrical
\cite{Lou03,Ger04} (typically for single layer dielectric beam
splitters) or non-symmetrical operator input-output operator
relations \cite{Man95,Gry10} (typically for cube beam-splitters).


A Mach-Zehnder interferometer (MZI) is a device composed of two
beam splitters and two mirrors \cite{Lou03}. Its versatility has
led to its use in countless experiments
\cite{Gra86,Rar90,Fra91,Kwi95}. Applying a single light quantum at
one input, the rate of photo-detection oscillates as the
path-length difference of the interferometer is swept
\cite{Gra86}. However, no coincidence counts are detected.
Applying pairs of light quanta at its inputs, specific
non-classical effects show up \cite{Rar90}.


The special interest in Quantum Optics stems from the fact that it
allowed a whole new set of \emph{Gedankenexperiments} to be
brought to reality (\emph{e.g.} quantum eraser
\cite{Scu82,Kim00,Wal02,Jac08} interaction-free measurements
\cite{EV93,Kwi95}, quantum-non-demolition (QND) experiments
\cite{Gle07}). A review of fundamental experiments in the field of
QO is given in Steinberg \emph{et al.} \cite{Ste96}.

The so-called ``semi-classical'' approach \cite{Man95,Gry10} in QO
explains many aspects of light (including the
photo-electric effect \cite{Man64}) however, only a full quantized
theory is able to distinguish between ``classical'' (\emph{e.g.}
coherent, thermal) and ``non-classical'' (\emph{e.g.} Fock, squeezed) states of light
\cite{Gra86,Asp91}. The Hanbury-Brown and Twiss \cite{Han57}
experiment that yielded the ``photon bunching'' effect for thermal
light is expected to show \emph{anti-bunching} for Fock states of
light, a completely non-classical effect. The first to prove the
existence of such non-classical states of light were Kimble,
Dagenais and Mandel \cite{Kim77}.


Using a beam splitter and a source of parametric down-conversion,
Hong, Ou and Mandel \cite{HOM87} experimentally proved the
existence of this non-classical effect for pairs of single-quantum
light states. Varying the arrival time of the ``single photon wave
packets'' they obtained what we now call the ``HOM dip'': a sharp
drop in the coincidence counts when the light quanta impinge
simultaneously on the beam splitter. Other variants of this
experiment exist, for example with independent light sources
\cite{Fea89,Rar05}.

The question if the localized ``photon wave packets'' impinging
simultaneously on the beam splitter tell the whole story arose
ever since the HOM experiment. In order to answer this question,
Pittman \emph{et al.} \cite{Pit96} performed the same HOM
interferometer experiment but with a voluntary time delay between
the ``photon wave packets'' at the beam splitter, compensated
thereafter before the detectors. The same dip was obtained in the
coincidence counting rate, thus proving that the reassuring image
of overlapping ``photon wave packets'' at the beam splitter is not
the key to this experiment and the authors conclude that ``the
intuitively comforting notion of the photons overlapping at the
beam splitter is not at the heart of the interference, but a mere
artifact of the particular geometry of the setups'' \cite{Pit96}.
Kim \emph{et al.} \cite{Kim99} and later Kim \cite{Kim03} went
even further, proving that separating the ``single photon pulses''
beyond the coherence time of the pump laser still yields the
famous HOM dip, with a visibility of more than $80$\%. The initial
experiment (without delays) was also performed with orthogonal
polarizations imposed to its input (\emph{i.e.} $\vert{V}\rangle$
and $\vert{H}\rangle$). As expected, no interference was found,
although the ``photon wave packets'' overlap at the beam splitter.
The author concludes that ``[$\ldots$] the photon bunching picture
often used in literature is indeed incorrect in general and should
not be used whenever possible''.

Bylander \emph{et al.} \cite{Byl03}, using single light quanta
from different sources in successive pulses show that the same
phenomenon takes place, if indistinguishability is assured.

It is often believed that only identical (\emph{i.e.} having the
same energy) single-quantum states of light can produce this type
of HOM interference. Indistinguishability \cite{Fey65} is indeed,
important, but on the detector side. Raymer \emph{et al.}
\cite{Ray10} propose an interference experiment of ``two photons
of different color''. Using an active beam splitter, the initially
distinguishable ``red'' and ``blue'' single-photon states can be
converted into indistinguishable ``green'' single-photon states,
hence the quantum interference leading to the HOM dip.

In spite of all these experiments, the localized ``photon wave
packet'' picture is widely found today in literature, giving the
impression that it is an unquestionable common place knowledge. We
can read such statements as ``the length of the photon wave
packet'', ``the photon wave packets overlap'' or ``the photons
were short'', and we can even see graphics depicting
Gaussian-damped sinusoids impinging on a beam splitter or on a
detector, thus giving the impression that the very localized
photon wave packet is taken more or less seriously.

We wish to add another argument in this paper in order to dispel
this simplistic view of light quanta impinging on a beam splitter.
However, compared to previous experiments, we propose a simple
setup, requiring only photon pair production (from a parametric
down-conversion source, for example), photo-detectors and regular
beam splitters. No polarization beam splitters and no expensive
pulse creation/selection equipment is required.

This paper is organized as follows. In Section
\ref{sec:QO_descrption_BS_MZI} we describe in the formalism of
Quantum Optics the beam-splitter and the Mach-Zehnder
interferometer. We discuss the anti-bunching effect of a
beam-splitter in Section \ref{sec:antibunching_BS}. The
quantum-optical description of a double MZI experiment is done in
Section \ref{sec:DOUBLE_MZI_QO_description}. Two experiments with
the double MZI experiment are proposed: in Section
\ref{sec:double_MZI_one_input_quantum} the experimental setup is
excited with a single-quantum Fock state while singles probability
counts are evaluated and in Section
\ref{sec:double_MZI_two_light_quanta} two single-quantum states
impinge simultaneously at its input while the main focus lies on
the coincidence photo-counts. Finally, conclusions are drawn in
Section \ref{sec:conclusions}.


\section{The quantum optical description of beam splitters and Mach-Zehnder interferometers}
\label{sec:QO_descrption_BS_MZI}

In the following we shall denote by $\hat{a}_k$
($\hat{a}_k^\dagger$) and, respectively,
$\hat{D}_k\left(\alpha\right)$, the annihilation (creation) and,
respectively, displacement operators acting at the port $k$. The
quantum state of light ${\vert\phi\rangle=\vert1_00_1\rangle}$
denotes a state with one quantum of light in mode (port) $0$ and
none in mode (port) $1$. Throughout this paper, all
photo-detectors are assumed to be ideal.

\subsection{The case of monochromatic light}
\label{subsec:QO_descrption_BS_MZI_monochromatic} For a lossless
beam splitter, the output annihilation operators ($\hat{a}_3$ and
$\hat{a}_2$) can be written in respect with the input field
operators \cite{Lou03} as
\begin{equation}
\label{eq:app:a3_fct_a0_a1} \hat{a}_3=T\hat{a}_1+R\hat{a}_0
\end{equation}
and
\begin{equation}
\label{eq:app:a2_fct_a0_a1} \hat{a}_2=R\hat{a}_1+T\hat{a}_0
\end{equation}
where $T$ and $R$ represent the transmission, and, respectively,
the reflection coefficients. The input field operators
($\hat{a}_0$ and $\hat{a}_1$) obey the usual commutation relations
$[\hat{a}_l,\hat{a}_k]=[\hat{a}_l^\dagger,\hat{a}_k^\dagger]=0$
and $[\hat{a}_l,\hat{a}_k^\dagger]=\delta_{lk}$ where
$\delta_{lk}$ is the Kronecker delta and $l,k=0,1$. We impose the
same commutation relations to the output field operators and end
up with the constraints
\begin{equation}
\vert{T}\vert^2+\vert{R}\vert^2=1
\end{equation}
and
\begin{equation}
RT^*+TR^*=0
\end{equation}
At this point we have the freedom to choose the phase of our
coefficients. When dealing with a balanced (50/50) beam splitter,
we shall use $T=1/\sqrt{2}$ and $R=i/\sqrt{2}$ \cite{Lou03,Ger04}.
From Eqs.~\eqref{eq:app:a3_fct_a0_a1} and
\eqref{eq:app:a2_fct_a0_a1} one can also obtain the ``inverse''
relations involving creation operators,
\begin{equation}
\label{eq:a0_dagger_in_respect_w_a2_a3_BS}
\hat{a}_0^\dagger=T\hat{a}_2^\dagger+R\hat{a}_3^\dagger
\end{equation}
and
\begin{equation}
\label{eq:a1_dagger_in_respect_w_a2_a3_BS}
\hat{a}_1^\dagger=R\hat{a}_2^\dagger+T\hat{a}_3^\dagger
\end{equation}
It is interesting to compare the output of a BS when at its input
one applies a single-quantum Fock state and, respectively, a
coherent state. In the first case we have
$\vert\psi_{in}\rangle=\hat{a}_1^\dagger\vert0\rangle$, therefore
\begin{equation}
\label{eq:BS_output_state_for_single_photon_in}
\vert\psi_{out}\rangle=R\vert1_20_3\rangle+T\vert0_21_3\rangle
\end{equation}
This is an \emph{entangled state} and detecting separately photons
in any of the output ports will simply yield a probability of
$\vert{R}\vert^2$ or $\vert{T}\vert^2$. However, the probability
of coincident counts at its output yields
\begin{equation}
P_c=\langle\psi_{out}\vert\hat{a}_2^\dagger\hat{a}_3^\dagger\hat{a}_3\hat{a}_2\vert\psi_{out}\rangle=0
\end{equation}
a result that is expected since we have a single quantum state
that cannot yield multiple detections. In the second case, we
describe the coherent state using the displacement operator
relation
$\hat{D}_1\left(\alpha\right)=\hat{D}_2\left(R\alpha\right)\hat{D}_3\left(T\alpha\right)$
\cite{Lou03,Ger04} and we find
\begin{equation}
\vert\psi_{out}\rangle=\hat{D}_2\left(R\alpha\right)\hat{D}_3\left(T\alpha\right)\vert0\rangle
=\vert\left(R\alpha\right)_2\left(T\alpha\right)_3\rangle
\end{equation}
yielding a \emph{non-entangled state}, fundamentally different
from a state given by
Eq.~\eqref{eq:BS_output_state_for_single_photon_in}. Even for very
small $\alpha$, such a state will yield a non-null coincidence
rate, as proven in the experiment of Aspect and Grangier
\cite{Asp91}.

Another interesting state at the input of a beam splitter would be
$\vert\psi_{in}\rangle=R\vert0_01_1\rangle+T\vert1_00_1\rangle$.
Using Eqs.~\eqref{eq:a0_dagger_in_respect_w_a2_a3_BS} and
\eqref{eq:a1_dagger_in_respect_w_a2_a3_BS}, we find the output
state
\begin{equation}
\label{eq:BS_output_state_for_entangled_input}
\vert\psi_{out}\rangle=2RT\vert1_20_3\rangle+(R^2+T^2)\vert0_21_3\rangle
\end{equation}
and in the case of a balanced beam splitter we have $R^2+T^2=0$,
therefore ${\vert\psi_{out}\rangle=\vert1_20_3\rangle}$. In this
case, the beam splitter transforms an \emph{entangled state} into
a \emph{non-entangled} one, where the light quantum always leaves
the beam splitter from the same port.

In the following (except when specifically stated) we will assume
\emph{balanced }($50/50$) beam splitters.

\begin{figure}
\centering
\includegraphics[width=1.8in]{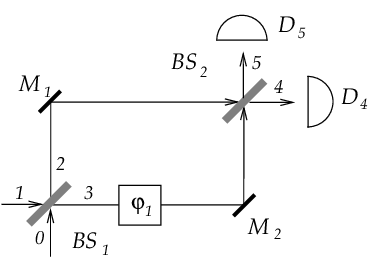}
\caption{A Mach-Zehnder interferometer. The delay $\varphi_1$
models the difference in optical path lengths between the two arms
of the interferometer. $D_4$ and $D_5$ denote the photo-detectors
placed at the two outputs of beam splitter $\text{BS}_2$.}
\label{fig:single_Mach_Zehnder_experiment}
\end{figure}

The  Mach-Zehnder interferometer (depicted in Fig.
\ref{fig:single_Mach_Zehnder_experiment}) is composed of two
mirrors and two beam splitters. The delay $\varphi_1$ models the
difference in optical path lengths between the two arms of the
interferometer. For monochromatic light quanta of frequency
$\omega$ we can write $\varphi_1=\omega\tau_1$ where $\tau_1$
denotes the time delay introduced. We can relate it to $z$, the
path length difference of the MZI and we obviously have
$\tau_1=z/c$ where $c$ is the speed of light in vacuum. The input
creation operators in respect with the output ones are obtained
from Eqs.~\eqref{eq:a0_dagger_in_respect_w_a2_a3_BS} and
\eqref{eq:a1_dagger_in_respect_w_a2_a3_BS}, applied to both beam
splitters. Combining them and considering the delay $\varphi_1$
applied to $\hat{a}_3^\dagger$, we end up with
\begin{equation}
\hat{a}_0^\dagger=-\sin\left(\varphi_1/2\right)\hat{a}_4^\dagger
+\cos\left(\varphi_1/2\right)\hat{a}_5^\dagger
\end{equation}
and
\begin{equation}
\hat{a}_1^\dagger=\cos\left(\varphi_1/2\right)\hat{a}_4^\dagger
+\sin\left(\varphi_1/2\right)\hat{a}_5^\dagger
\end{equation}
A single-quantum Fock state
$\vert\psi_{in}\rangle=\vert0_01_1\rangle$ applied to the MZI is
transformed into
\begin{equation}
\vert\psi_{out}\rangle=\cos\left(\varphi_1/2\right)\vert0_51_4\rangle
+\sin\left(\varphi_1/2\right)\vert0_41_5\rangle
\end{equation}
yielding the well-known sine-like probability of singles
detection,
\begin{equation}
\label{eq:P4_MZI_10_input_monochromatic}
P_4=\vert\langle1_40_5\vert\psi_{out}\rangle\vert^2
=\frac{1}{2}\Big(1+\cos\left(\omega\tau_1\right)\Big)
\end{equation}
and 
\begin{equation}
\label{eq:P5_MZI_10_input_monochromatic}
P_5=\vert\langle0_41_5\vert\psi_{out}\rangle\vert^2
=\frac{1}{2}\Big(1-\cos\left(\omega\tau_1\right)\Big)
\end{equation}
in respect with $\varphi_1=\omega\tau_1$. A more interesting situation appears
if we apply two simultaneously impinging light quanta on the first
beam splitter \emph{i.e}
$\vert\psi_{in}\rangle=\vert1_01_1\rangle$. This time, we have the
output state
\begin{eqnarray}
\vert\psi_{out}\rangle=-1/\sqrt{2}\sin\left(\varphi_1\right)\vert2_40_5\rangle
\qquad\qquad\qquad
\nonumber\\
+1/\sqrt{2}\sin\left(\varphi_1\right)\vert0_42_5\rangle
+\cos\left(\varphi_1\right)\vert1_41_5\rangle
\end{eqnarray}
The probability of coincident detections at $D_4$ and $D_5$ is
given by
\begin{equation}
\label{eq:Pc_MZI_11_input_monochromatic}
P_c=\vert\langle1_41_5\vert\psi_{out}\rangle\vert^2
=\frac{1}{2}\Big(1+\cos\left(2\varphi_1\right)\Big)
\end{equation}
showing that indeed, by continuously varying $\varphi_1$ the
coincidence probability $P_c$ goes back and forth between $0$ to
$1$. For $\varphi_1=0$ we have
$\vert1_01_1\rangle\rightarrow\vert1_41_5\rangle$, or, in other
words, the MZI is \emph{transparent} to the $\vert1_01_1\rangle$
input state.

It is worthwhile to note that in
Eq.~\eqref{eq:Pc_MZI_11_input_monochromatic} the frequency of
variation of the interference fringes is twice the frequency of
the input light source. This phenomenon was experimentally tested
by Rarity \emph{et al.} \cite{Rar90}. They used pairs of
correlated light quanta from a $\lambda=826.8$ nm source obtained
from down-converting a 413.4 nm krypton-ion laser. They found that
the spatial period of the probability coincidence corresponds to
the $413$ nm of the pump laser.

\subsection{The case of non-monochromatic light}
\label{subsec:QO_descrption_BS_MZI_nonmonochromatic} Extension to
a continuum of modes has been already considered
\cite{Lou03,Blo90}, often for two-photon states impinging on a
beam splitter \cite{Fea89,Cam90}. Assuming non-monochromatic but
\emph{narrowband} light quanta, Legero \emph{et al.}
\cite{Leg03,Leg06} extend the result from
Eq.~\eqref{eq:BS_antibunching} with a space-time domain
description by considering spatio-temporal mode functions
$\zeta_l\left(z,t\right)=\epsilon_l\left(t-z/c\right)\text{e}^{-i\phi_l\left(t-z/c\right)}$
of a single-mode input radiation (input ports are labelled $l=0$
and $l=1$). By placing the beam-splitter at $z=0$ the space
coordinate $z$ can be omitted. The mode functions are assumed to
be normalized, so that
$\int{\vert\epsilon_l\left(t\right)\vert^2\text{d}t}=1$. Then, the
input electric field operators can be written as
$\hat{E}_l^{(+)}\left(t\right)=\zeta_l\left(t\right)\hat{a}_l$ and
$\hat{E}_l^{(-)}\left(t\right)=\zeta_l^*\left(t\right)\hat{a}_l^\dagger$
with $l=0,1$ while the output electric fields are
\begin{equation}
\label{eq:E_2_plus_fct_zet_0_zeta_1}
\hat{E}_2^{(+)}\left(t\right)=\frac{1}{\sqrt{2}}\left(\zeta_0\left(t\right)\hat{a}_0+i\zeta_1\left(t\right)\hat{a}_1\right)
\end{equation}
and
\begin{equation}
\label{eq:E_3_plus_fct_zet_0_zeta_1}
\hat{E}_3^{(+)}\left(t\right)=\frac{1}{\sqrt{2}}\left(i\zeta_0\left(t\right)\hat{a}_0+\zeta_1\left(t\right)\hat{a}_1\right)
\end{equation}
The output field operators for the MZI are
\begin{equation}
\label{eq:E_4_plus_fct_zeta_0_zeta_1}
\hat{E}_4^{(+)}\left(t\right)=\frac{\zeta_0\left(t\right)-\zeta_0\left(t-\tau_1\right)}{2}\hat{a}_0
+\frac{\zeta_1\left(t\right)+\zeta_1\left(t-\tau_1\right)}{2}i\hat{a}_1
\end{equation}
and
\begin{equation}
\label{eq:E_5_plus_fct_zet_0_zeta_1}
\hat{E}_5^{(+)}\left(t\right)=\frac{\zeta_0\left(t\right)+\zeta_0\left(t-\tau_1\right)}{2}i\hat{a}_0
-\frac{\zeta_1\left(t\right)-\zeta_1\left(t-\tau_1\right)}{2}\hat{a}_1
\end{equation}
If the input state $\vert\psi_{in}\rangle=\vert0_01_1\rangle$ is
applied to our MZI, we find the probability of singles detection
at $D_4$ given by
\begin{eqnarray}
P_4=\langle\psi_{in}\vert\hat{E}_4^{(-)}\left(t_0\right)\hat{E}_4^{(+)}\left(t_0\right)\vert\psi_{in}\rangle
\qquad\qquad
\nonumber\\ 
\qquad\qquad
=\frac{1}{4}\big\vert\zeta_1\left(t_0\right)+\zeta_1\left(t_0-\tau_1\right)\big\vert^2
\end{eqnarray}
If we consider a Gaussian space-time mode function
Eq.~\eqref{eq:zeta1_Gaussian} for $\zeta_1\left(t\right)$, after
$t_0$-integration (a similar -- slightly more complicated --
computation is done in Appendix
\ref{sec:app:P6_single_photon_detection_2MZI}) we arrive at
\begin{equation}
\label{eq:P4_MZI_10_input_non_monochromatic}
P_4
=\frac{1}{2}\left(1+\text{e}^{-\frac{\tau_1^2}{\sigma^2}}\cos\left(\omega\tau_1\right)\right)
\end{equation}
where this time $\omega$ represents the central frequency of our narrowband light quantum.
We find the same oscillatory behavior from
Eq.~\eqref{eq:P4_MZI_10_input_monochromatic}, however damped
because of the finite bandwidth assumed for our light quantum. If
we input the state $\vert\psi_{in}\rangle=\vert1_01_1\rangle$ to
the MZI, one finds the probability of coincident counts
\begin{eqnarray}
\label{eq:Prob_coinc_single_MZI}
P_c\left(t_0,\tau,\tau_d\right)=\langle\psi_{in}\vert\hat{E}_4^{(-)}\left(t_0\right)\hat{E}_5^{(-)}\left(t_0+\tau_d\right)
\qquad
\nonumber\\ 
\qquad
\hat{E}_5^{(+)}\left(t_0+\tau_d\right)\hat{E}_4^{(+)}\left(t_0\right)\vert\psi_{in}\rangle
\end{eqnarray}
where $\tau_d$ corresponds to the time delay between the
detections at $D_4$ and $D_5$, related to their distance from the
beam splitter $\text{BS}_2$. If we set $\tau_d=0$ in
Eq.~\eqref{eq:Prob_coinc_single_MZI} and replace the functions
$\zeta_0\left(t\right)$ and $\zeta_1\left(t\right)$ with the
Gaussian expressions from Eqs.~\eqref{eq:zeta0_Gaussian} and
\eqref{eq:zeta1_Gaussian}, after time integration one gets
\begin{equation}
\label{eq:Pc_MZI_11_input_non_monochromatic}
P_c=\frac{1}{2}\left(1+\text{e}^{-\frac{\tau_1^2}{\sigma^2}}\cos\left(2\omega\tau_1\right)\right)
\end{equation}
where we find again the same oscillatory behavior from
Eq.~\eqref{eq:Pc_MZI_11_input_monochromatic}, damped however by an
exponential factor.

\section{The anti-bunching effect on a beam splitter}
\label{sec:antibunching_BS}

\subsection{The case of monochromatic light}
\label{subsec:antibunching_BS_monochromatic} We consider the input
state $\vert\psi\rangle_{in}=\vert1_01_1\rangle$ \emph{i.e.} two
simultaneously impinging light quanta on a beam splitter with
transmission (reflection) coefficient $T$ ($R$). Using
Eqs.~\eqref{eq:a0_dagger_in_respect_w_a2_a3_BS} and
\eqref{eq:a1_dagger_in_respect_w_a2_a3_BS}, we obtain the output
state
\begin{eqnarray}
\label{eq:BS_antibunching}
\vert\psi_{out}\rangle=\sqrt{2}\left(RT\vert0_22_3\rangle+RT\vert2_20_3\rangle\right)
\qquad
\nonumber\\ 
\qquad
+(R^2+T^2)\vert1_21_3\rangle
\end{eqnarray}
If the beam-splitter is balanced, we have $R^2+T^2=0$, therefore
the $\vert1_21_3\rangle$ output state from
Eq.~\eqref{eq:BS_antibunching} vanishes. In other words, both
light quanta will always exit the beam splitter through the same
port. This is the \emph{antibunching} or HOM effect.

\subsection{The case of non-monochromatic light}
\label{subsec:antibunching_BS_nonmonochromatic} The input state is
still assumed to be
$\vert\psi_{in}\rangle=\vert1_01_1\rangle$. The non-monochromatic
character of the light quanta will be modelled through the field
operators $\hat{E}_2^{(+)}\left(t\right)$ and
$\hat{E}_3^{(+)}\left(t\right)$. The probability of joint
detection at the outputs of the beam splitter at times $t_0$ and,
respectively $t_0+\tau_d$ is
\begin{eqnarray}
P_c\left(t_0,\tau_d\right)=\langle\psi_{in}\vert\hat{E}_2^{(-)}\left(t_0\right)\hat{E}_3^{(-)}\left(t_0+\tau_d\right)
\qquad
\nonumber\\ 
\qquad
\hat{E}_3^{(+)}\left(t_0+\tau_d\right)\hat{E}_2^{(+)}\left(t_0\right)\vert\psi_{in}\rangle
\end{eqnarray}
and after a series of calculations \cite{Leg03,Leg06} the final
result reads
\begin{equation}
\label{eq:NBS_HOM_dip_spatiotemporal}
P_c\left(t_0,\tau_d\right)=\frac{1}{4}\big\vert\zeta_0\left(t_0+\tau_d\right)\zeta_1\left(t_0\right)
-\zeta_0\left(t_0\right)\zeta_1\left(t_0+\tau_d\right)\big\vert^2
\end{equation}
implying that indeed, for simultaneously impinging light quanta on
the beam splitter, detected at the same time ($\tau_d=0$) , the
probability of coincident counts is $P_c=0$, regardless of the
temporal shapes of $\zeta_0\left(t\right)$ and
$\zeta_1\left(t\right)$. We can take Gaussian spatio-temporal mode
functions,
\begin{equation}
\label{eq:zeta0_Gaussian}
\zeta_{0}\left(t\right)=\left(\frac{2}{\pi\sigma^2}\right)^{\frac{1}{4}}\text{e}^{-\frac{(t-\tau_e/2)^2}{\sigma^2}-i\omega{t}}
\end{equation}
and
\begin{equation}
\label{eq:zeta1_Gaussian}
\zeta_{1}\left(t\right)=\left(\frac{2}{\pi\sigma^2}\right)^{\frac{1}{4}}\text{e}^{-\frac{(t+\tau_e/2)^2}{\sigma^2}-i\omega{t}}
\end{equation}
where $\sigma$ quantifies the time spread of our light quanta and
$\tau_e$ can be seen as the time difference of the impinging
Gaussian ``photon wave packets'' on the beam splitter. The
probability of coincident counts from
Eq.~\eqref{eq:NBS_HOM_dip_spatiotemporal} is easily evaluated to
\begin{equation}
P_c\left(t_0,\tau_e,\tau_d\right)
=\frac{\cosh\left(2\tau_d\tau_e/\sigma^2\right)-1}{\pi\sigma^2}\text{e}^{-\frac{4t_0(t_0+\tau_d)+\tau_e^2+2\tau_d^2}{\sigma^2}}
\end{equation}
and integrating it over all possible values of $t_0$ we get
\begin{equation}
\label{eq:NBS_HOM_dip_Gaussians_integrated_over_t0}
P_c\left(\tau_e,\tau_d\right)=\frac{\cosh\left(2\tau_d\tau_e/\sigma^2\right)-1}{2\sqrt{\pi}\sigma}\text{e}^{-\frac{\tau_e^2+\tau_d^2}{\sigma^2}}
\end{equation}
If the photo-detectors are ``slow'', we integrate
$P_c\left(\tau_e,\tau_d\right)$ over the detection time difference
$\tau_d$, yielding
\begin{equation}
\label{eq:NBS_HOM_dip_Gaussians_integrated_over_t0_and_over_tau_d}
P_c\left(\tau_e\right)=\frac{1}{2}\left(1-\text{e}^{-\frac{\tau_e^2}{\sigma^2}}\right)
\end{equation}
This detection probability shows indeed the very famous ``HOM
dip'', experimentally measured by Hong, Ou and Mandel
\cite{HOM87}, where they report a time measurement of a
``sub-picosecond photon wave packet''. The fact that those
``photon wave packets'' meet at the beam splitter seems to cause
the famous dip in the rate of coincidences at the detector.
Eq.~\eqref{eq:NBS_HOM_dip_Gaussians_integrated_over_t0_and_over_tau_d}
quantifies this result, implying that a null coincidence
probability can only be achieved if $\tau_e=0$ \emph{i.e.} if both
``photon wave packets'' are impinging simultaneously on the beam
splitter.

\section{The quantum optical description of a double MZI experiment}
\label{sec:DOUBLE_MZI_QO_description} 
\begin{figure}
\centering
\includegraphics[width=2.5in]{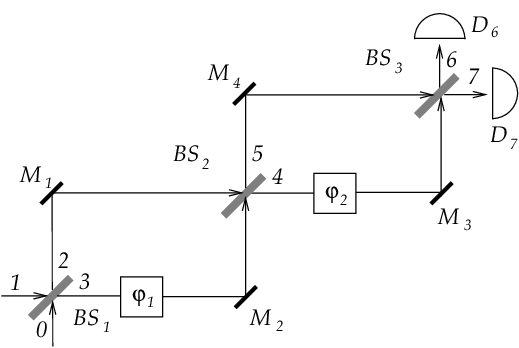}
\caption{The double MZI experiment proposed in Section
\ref{sec:DOUBLE_MZI_QO_description}. The first MZI is composed of
the beam splitters $\text{BS}_1$ and $\text{BS}_2$ together with
the mirrors $\text{M}_1$ and $\text{M}_2$. Similarly, the second
MZI is composed of $\text{BS}_2$ , $\text{BS}_3$, $\text{M}_3$ and
$\text{M}_4$. $\varphi_1$ and, respectively $\varphi_2$ represent
delays voluntarily introduced in the paths labelled ``3'' and,
respectively, ``4''.} \label{fig:double_Mach_Zehnder_experiment}
\end{figure}

In the following, we
introduce and discuss a double MZI interferometer experiment
(depicted in Fig. \ref{fig:double_Mach_Zehnder_experiment}). The
beam splitter $\text{BS}_1$, together with the beam splitter
$\text{BS}_2$ and the two mirrors $\text{M}_1$ and $\text{M}_2$
form the first MZI. In the lower path a delay $\varphi_1$ is
introduced. The second MZI is composed of the beam splitters
$\text{BS}_2$ and $\text{BS}_3$, together with the two
corresponding mirrors $\text{M}_3$ and $\text{M}_4$. In the lower
path of this interferometer a delay $\varphi_2$ is present. It is
assumed that, with the corresponding delays taken out of the
experiment, each MZI has equal length arms. Photo-detectors $D_6$
and $D_7$ are installed at the two outputs of $\text{BS}_3$.

\subsection{The case of monochromatic light}
\label{subsec:DOUBLE_MZI_QO_description_monochromatic} The input
(creation) field operators (ignoring some common phase factors) in
respect with the output field operators are (see Appendix
\ref{sec:app:compute_a0_a1_dagger})
\begin{eqnarray}
\label{eq:a0_dagger_in_respect_w_a6_a7} \hat{a}_0^\dagger
=\frac{-i\sin\left(\frac{\varphi_2}{2}\right)
-\text{e}^{-i\varphi_1}\cos\left(\frac{\varphi_2}{2}\right)}{\sqrt{2}}
\hat{a}_6^\dagger
\qquad\qquad  
\nonumber\\ 
\qquad\qquad 
+\frac{i\cos\left(\frac{\varphi_2}{2}\right)-\text{e}^{-i\varphi_1}\sin\left(\frac{\varphi_2}{2}\right)}{\sqrt{2}}\hat{a}_7^\dagger
\end{eqnarray}
and
\begin{eqnarray}
\label{eq:a1_dagger_in_respect_w_a6_a7} \hat{a}_1^\dagger
=\frac{\sin\left(\frac{\varphi_2}{2}\right)+i\text{e}^{-i\varphi_1}\cos\left(\frac{\varphi_2}{2}\right)}{\sqrt{2}}\hat{a}_6^\dagger
\qquad\qquad  
\nonumber\\ 
\qquad\qquad 
+\frac{-\cos\left(\frac{\varphi_2}{2}\right)+i\text{e}^{-i\varphi_1}\sin\left(\frac{\varphi_2}{2}\right)}{\sqrt{2}}\hat{a}_7^\dagger
\end{eqnarray}
These relations will be used in the next sections in order to
compute the output state of our system.

\subsection{The case of non-monochromatic light}
\label{subsec:DOUBLE_MZI_QO_description_nonmonochromatic} The
output field operators $\hat{E}_6^{(+)}\left(t\right)$ and
$\hat{E}_7^{(+)}\left(t\right)$ expressed in respect with the
input fields are (see details in Appendix
\ref{sec:app:compute_E6_E7_plus})
\begin{eqnarray}
\label{eq:E6__non_monochrom_tau_e_tau_d}
\hat{E}_6^{(+)}\left(t\right) =\frac{1}{2\sqrt{2}}\Big(\hat{a}_0
\big(\zeta_0\left(t-\tau_2\right)-\zeta_0\left(t-\tau_1-\tau_2\right)
\quad 
\nonumber\\ 
-\zeta_0\left(t\right)-\zeta_0\left(t-\tau_1\right)\big)
+i\hat{a}_1 \big(\zeta_1\left(t-\tau_2\right)
\nonumber\\ 
+\zeta_1\left(t-\tau_1-\tau_2\right)-\zeta_1\left(t\right)+\zeta_1\left(t-\tau_1\right)\big)\Big)
\end{eqnarray}
and
\begin{eqnarray}
\label{eq:E7__non_monochrom_tau_e_tau_d}
\hat{E}_7^{(+)}\left(t\right) =\frac{1}{2\sqrt{2}}\Big(i\hat{a}_0
\big(\zeta_0\left(t-\tau_2\right)-\zeta_0\left(t-\tau_1-\tau_2\right)
\quad 
\nonumber\\ 
+\zeta_0\left(t\right)+\zeta_0\left(t-\tau_1\right)\big)
+\hat{a}_1 \big(-\zeta_1\left(t-\tau_2\right)
\nonumber\\ 
-\zeta_1\left(t-\tau_1-\tau_2\right)-\zeta_1\left(t\right)+\zeta_1\left(t-\tau_1\right)\big)\Big)
\end{eqnarray}
We obviously have
$\hat{E}_6^{(-)}\left(t\right)=[\hat{E}_6^{(+)}\left(t\right)]^\dagger$
and
$\hat{E}_7^{(-)}\left(t\right)=[\hat{E}_7^{(+)}\left(t\right)]^\dagger$.
These field operators will be used in the next sections in order
to find singles and coincidence photo-detection probabilities.

\section{The double MZI with a single light quantum at one input}
\label{sec:double_MZI_one_input_quantum} In this section we
analyze the double MZI experimental setup when at its input we
have a single quantum (monochromatic or non-monochromatic) Fock
state.

\subsection{The case of monochromatic light}
\label{subsec:double_MZI_one_input_quantum_monochromatic} The
input state can be written as
$\vert\psi_{in}\rangle=\hat{a}_1^\dagger\vert0\rangle$ and taking
into account Eq.~\eqref{eq:a1_dagger_in_respect_w_a6_a7} we have
\begin{eqnarray}
\label{eq:psi_out_2MZI_10_input_state}
\vert\psi_{out}\rangle=\frac{\sin\left(\frac{\varphi_2}{2}\right)
+i\text{e}^{-i\varphi_1}\cos\left(\frac{\varphi_2}{2}\right)}{\sqrt{2}}\vert1_60_7\rangle
\qquad
\nonumber\\ 
\qquad+\frac{-\cos\left(\frac{\varphi_2}{2}\right)
+i\text{e}^{-i\varphi_1}\sin\left(\frac{\varphi_2}{2}\right)}{\sqrt{2}}\vert0_61_7\rangle
\end{eqnarray}
The probability of single-photon detection at the detector $D_6$
can be easily computed, yielding
\begin{equation}
\label{eq:P6_2MZI_10_input_state}
P_6
=\vert\langle1_60_7\vert\psi_{out}\rangle\vert^2
=\frac{1}{2}\Big(1+\sin(\varphi_1)\sin(\varphi_2)\Big)
\end{equation}
Similarly, computing the single-photon detection probability at
the detector $D_7$ we find
\begin{equation}
\label{eq:P7_2MZI_10_input_state}
P_7
=\vert\langle0_61_7\vert\psi_{out}\rangle\vert^2
=\frac{1}{2}\Big(1-\sin(\varphi_1)\sin(\varphi_2)\Big)
\end{equation}
and we have $P_6+P_7=1$, as expected. However,
Eqs.~\eqref{eq:P6_2MZI_10_input_state} and
\eqref{eq:P7_2MZI_10_input_state} imply that the sine-like
variation of the detection probabilities in respect with the
length difference of the two arms for a single MZI (as found in
Section \ref{subsec:QO_descrption_BS_MZI_monochromatic}) is no
longer true. Indeed, by setting for example $\sin(\varphi_1)=0$,
we end up with $P_6=P_7=1/2$, no matter what value $\varphi_2$
takes. The same is true for $\varphi_1$ if we fix
$\sin(\varphi_2)=0$.

This apparent paradox of Eqs.~\eqref{eq:P6_2MZI_10_input_state}
and \eqref{eq:P7_2MZI_10_input_state} can be explained if we
consider, for example, the state of the field at the output of
beam splitter $\text{BS}_2$ for $\sin\varphi_1=0$. Indeed, using
Eq.~\eqref{eq:BS_output_state_for_entangled_input} we find
$\vert\psi_{45}\rangle=\vert0_51_4\rangle$. In other words, our
light quantum \emph{always takes only one arm} in the second MZI.
The delay $\varphi_2$ becomes therefore, useless.

If we set $\varphi_1$ so that we have $\sin(\varphi_1)=\beta$, the
probability of single-photon detection at the detector $D_6$ will
be $P_6=1/2\left(1+\beta\sin(\varphi_2)\right)$, in other words,
the delay $\varphi_1$ modulates the photo-detection probability
$P_6\left(\varphi_2\right)$.

The probability of coincident counts is
$P_c=0$, an expected result since a single light quantum cannot
yield multiple output detections. However if we apply a coherent
source instead of the Fock state, the situation dramatically
changes. Indeed, for a state
$\vert\psi_{in}\rangle=\vert0_0\alpha_1\rangle$ we find
\begin{eqnarray}
\vert\psi_{out}\rangle=\hat{D}_6\left(\alpha\frac{\sin\left(\varphi_2/2\right)+i\text{e}^{-i\varphi_1}\cos\left(\varphi_2/2\right)}{\sqrt{2}}\right)
\nonumber\\
\hat{D}_7\left(\alpha\frac{-\cos\left(\frac{\varphi_2}{2}\right)+i\text{e}^{-i\varphi_1}\sin\left(\frac{\varphi_2}{2}\right)}{\sqrt{2}}\right)
\vert0\rangle
\end{eqnarray}
The rate of coincidence detection is proportional to
\begin{equation}
N_c\sim
\frac{\vert\alpha\vert^4}{4}\Big(1-\sin^2\left(\varphi_1\right)\sin^2\left(\varphi_2\right)\Big)
\end{equation}
For the singles rates one easily finds
\begin{equation}
N_6\sim\frac{\vert\alpha\vert^2}{2}\Big(1+\sin\left(\varphi_1\right)\sin\left(\varphi_2\right)\Big)
\end{equation}
and
\begin{equation}
N_7\sim\frac{\vert\alpha\vert^2}{2}\Big(1-\sin\left(\varphi_1\right)\sin\left(\varphi_2\right)\Big)
\end{equation}
implying $N_c/N_6N_7=1$ \emph{i.e.} we have \emph{no antibunching}
with a coherent state.

\subsection{The case of non-monochromatic light}
\label{subsec:double_MZI_one_input_quantum_nonmonochromatic} 
\begin{figure}
\centering
\includegraphics[width=3.5in]{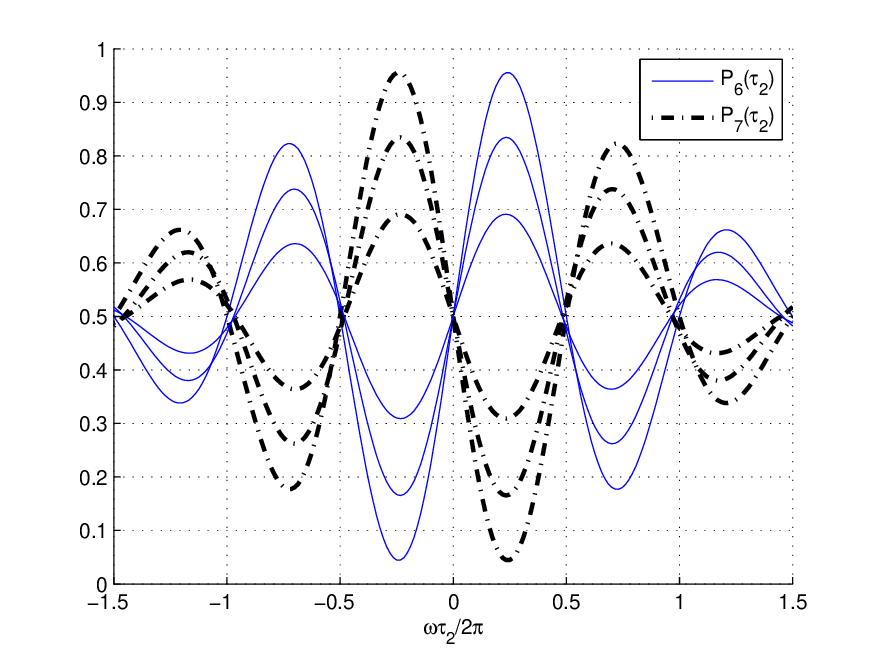}
\caption{The probability of single-photon detection for
$P_6\left(\tau_2\right)$ and, respectively,
$P_7\left(\tau_2\right)$ given by
Eq.~\eqref{eq:P6_2MZI_non_monochrom_tau1_fixed} and, respectively,
Eq.~\eqref{eq:P7_2MZI_non_monochrom_tau1_fixed} for $\beta=1$
(highest amplitude curves), $\beta=0.707$ and $\beta=0.4$ (lowest
amplitude curves). We used for this simulation $\sigma\omega=5$.}
\label{fig:MATLAB_P6_P7_versus_tau2_fixed_tau1_2MZI}
\end{figure}

If we
suppose a single light quantum having some spatio-temporal
extension $\zeta_1\left(t\right)$, we characterize the probability
of single-photon detection at time $t_0$ at the detector $D_6$ by
\begin{equation}
P_6=\langle\psi_{in}\vert\hat{E}^-_6\left(t_0\right)\hat{E}^+_6\left(t_0\right)\vert\psi_{in}\rangle
\end{equation}
Using Eq.~\eqref{eq:E6__non_monochrom_tau_e_tau_d}, replacing the
Gaussian waveforms and time-integrating the result (see details in
Appendix \ref{sec:app:P6_single_photon_detection_2MZI}) takes us
to
\begin{eqnarray}
\label{eq:P6_2MZI_non_monochrom_10_input_state}
P_6=\frac{1}{4}\bigg(2+\text{e}^{-\frac{(\tau_1-\tau_2)^2}{2\sigma^2}}\cos\left(\omega(\tau_1-\tau_2)\right)
\qquad\qquad 
\nonumber\\
\qquad 
-\text{e}^{-\frac{(\tau_1+\tau_2)^2}{2\sigma^2}}\cos\left(\omega(\tau_1+\tau_2)\right)\bigg)
\end{eqnarray}
Performing the same computations for the detector $D_7$ yields
\begin{eqnarray}
\label{eq:P7_2MZI_non_monochrom_10_input_state}
P_7=\frac{1}{4}\bigg(2-\text{e}^{-\frac{(\tau_1-\tau_2)^2}{2\sigma^2}}\cos\left(\omega(\tau_1-\tau_2)\right)
\qquad\qquad 
\nonumber\\
\qquad 
+\text{e}^{-\frac{(\tau_1+\tau_2)^2}{2\sigma^2}}\cos\left(\omega(\tau_1+\tau_2)\right)\bigg)
\end{eqnarray}
If we fix $\tau_1$ so that $\sin\left(\omega\tau_1\right)=\beta$
and denoting $\kappa=\tau_1/\sigma^2$,
$\gamma=\sqrt{1-\beta^2}\text{e}^{-\tau_1^2/2\sigma^2}$ and
$\delta=\beta\text{e}^{-\tau_1^2/2\sigma^2}$, we can rewrite the
singles detection probabilities as
\begin{eqnarray}
\label{eq:P6_2MZI_non_monochrom_tau1_fixed}
P_6=\frac{1}{2}\bigg(1-\gamma\text{e}^{-\frac{\tau_2^2}{2\sigma^2}}
\sinh\left(\kappa\tau_2\right)\cos\left(\omega\tau_2\right)
\qquad\qquad 
\nonumber\\
\qquad 
-\delta\text{e}^{-\frac{\tau_2^2}{2\sigma^2}}
\cosh\left(\kappa\tau_2\right)\sin\left(\omega\tau_2\right)\bigg)
\end{eqnarray}
and 
\begin{eqnarray}
\label{eq:P7_2MZI_non_monochrom_tau1_fixed}
P_7=\frac{1}{2}\bigg(1+\gamma\text{e}^{-\frac{\tau_2^2}{2\sigma^2}}
\sinh\left(\kappa\tau_2\right)\cos\left(\omega\tau_2\right)
\qquad\qquad 
\nonumber\\
\qquad 
+\delta\text{e}^{-\frac{\tau_2^2}{2\sigma^2}}
\cosh\left(\kappa\tau_2\right)\sin\left(\omega\tau_2\right)\bigg)
\end{eqnarray}
We depict in Fig.
\ref{fig:MATLAB_P6_P7_versus_tau2_fixed_tau1_2MZI} $P_6$ and $P_7$
for three different values of $\beta$. This parameter $\beta$
imposes constraints on the maximum amplitude of the sine-like
behavior of the probabilities $P_6$ and $P_7$ while the Gaussian
shaping causes them to ``fade'' towards the value of $1/2$ as
$\tau_2$ increases. For $\beta=0$ we obviously have
$P_6\left(\tau_2\right)=P_7\left(\tau_2\right)=1/2$, whatever the
value of $\tau_2$.

If we consider the transition to monochromatic light quanta
(\emph{i.e.} when $\sigma\to\infty$), it is easy to show that we
obtain again $P_6$ and, respectively, $P_7$ given by
Eqs.~\eqref{eq:P6_2MZI_10_input_state} and, respectively,
\eqref{eq:P7_2MZI_10_input_state}.

\section{The double MZI with two simultaneously impinging light quanta at its inputs}
\label{sec:double_MZI_two_light_quanta} In this section we
describe an experiment with the double MZI depicted Fig.
\ref{fig:double_Mach_Zehnder_experiment}, able to show that the
antibunching effect on a beam splitter has nothing to do with
``photon wave packets meeting at the beam splitter''. Both the
monochromatic and the non-monochromatic cases are discussed.

\subsection{The case of monochromatic light}
\label{subsec:double_MZI_two_light_quanta_monochromatic} With
arbitrary delays $\varphi_1$ and $\varphi_2$, for the input state
$\vert\psi_{in}\rangle=\vert1_01_1\rangle$ 
by using Eqs.~\eqref{eq:a0_dagger_in_respect_w_a6_a7} and
\eqref{eq:a1_dagger_in_respect_w_a6_a7} one finds the output state
\begin{eqnarray}
\label{eq:double_MZI_output_state}
\vert\psi_{out}\rangle=\frac{\sin\left(\varphi_1\right)\cos\left(\varphi_2\right)
-i\cos\left(\varphi_1\right)}{\sqrt{2}}\vert0_62_7\rangle
\nonumber\\
-\frac{\sin\left(\varphi_1\right)\cos\left(\varphi_2\right)-i\cos\left(\varphi_1\right)}{\sqrt{2}}\vert2_60_7\rangle
\nonumber\\
+\sin\left(\varphi_1\right)\sin\left(\varphi_2\right)\vert1_61_7\rangle
\end{eqnarray}
The probability of coincident detections at the outputs detectors
$D_6$ and $D_7$ is given by
\begin{equation}
\label{eq:double_MZI_Pc_monochromatic}
P_c\left(\varphi_1,\varphi_2\right)
=\vert\langle1_61_7\vert\psi_{out}\rangle\vert^2
=\sin^2\left(\varphi_1\right)\sin^2\left(\varphi_2\right)
\end{equation}
If we take out both delays from the experimental setup, we end up
with the same type of transformation given by
Eq.~\eqref{eq:BS_antibunching} \emph{i.e.} the whole experiment is
equivalent to a \emph{beam splitter}. At this point we can claim
that the ``photon wave packets'' meet at the beam splitter
$\text{BS}_2$. However, the same transformation can be obtained
with the delays in place, by choosing $\varphi_1=\varphi_2=m\pi$
where $m\in\mathbb{N}^*$. These delays can be made arbitrarily
large, the only limitation being imposed by the coherence
properties of the light source. This way, the output state from
\eqref{eq:double_MZI_output_state} becomes
\begin{eqnarray}
\label{eq:double_MZI_output_state_antiBunch}
\vert\psi_{out}\rangle=-\frac{i}{\sqrt{2}}\vert0_62_7\rangle
+\frac{i}{\sqrt{2}}\vert2_60_7\rangle
\end{eqnarray}
\emph{i.e.} a perfect anti-correlation that we expect to show the
famous ``HOM dip'' in the case of non-monochromatic light quanta.
However, this time we have no possible way of having ``overlapping
photon wave packets'' at the beam splitter\footnote{Strictly
speaking, this experiment performs exactly the opposite of
Eq.~\eqref{eq:BS_antibunching}, namely we have
$i/\sqrt{2}\vert0_22_3\rangle-i/\sqrt{2}\vert2_20_3\rangle\rightarrow\vert1_41_5\rangle$,
which is, of course, perfectly equivalent to
Eq.~\eqref{eq:BS_antibunching}. But if we insist on finding
exactly the anti-bunching effect from
Eq.~\eqref{eq:BS_antibunching} at $\text{BS}_2$, all we have to do
is add another MZI in front of the first one.} $\text{BS}_2$.

This experiment can also be seen as a delayed ``photon wave
packets'' HOM interferometer at the beam splitter $\text{BS}_2$
while the two other beam splitters act as quantum erasers, so that
the interfering paths at $\text{BS}_2$ become indiscernible.

\subsection{The case of non-monochromatic light}
\label{subsec:double_MZI_two_light_quanta_nonmonochromatic}
Extending the above results to spatio-temporal modes, the
probability of coincident counts at detectors $D_6$ and $D_7$, at
times $t_0$ and, respectively, $t_0+\tau_d$ is given by
\begin{eqnarray}
\label{eq:Prob_coinc_double_MZI_01}
P_c\left(t_0,\tau_1,\tau_2,\tau_d\right)=\langle\psi_{in}\vert\hat{E}_6^{(-)}\left(t_0\right)\hat{E}_7^{(-)}\left(t_0+\tau_d\right)
\nonumber\\ 
\hat{E}_7^{(+)}\left(t_0+\tau_d\right)\hat{E}_6^{(+)}\left(t_0\right)\vert\psi_{in}\rangle
\end{eqnarray}
We will be interested in the time-integrated detection probability
over all $t_0$ and, eventually, for the non time-resolved
detection, integrated over $\tau_d$, also.

\begin{figure}
\centering
\includegraphics[width=3.5in]{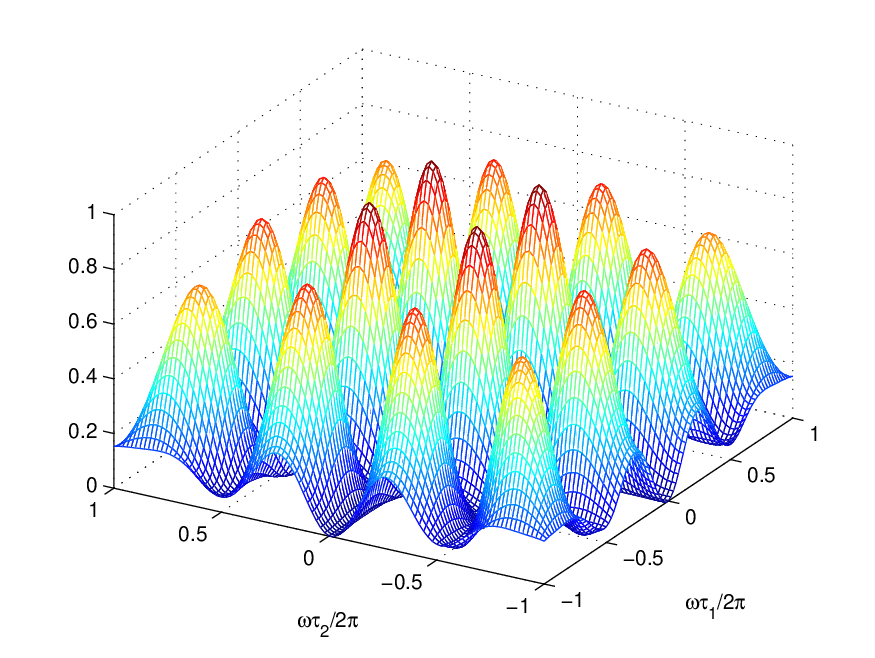}
\caption{The probability of coincident counts
$P_c\left(\tau_1,\tau_2\right)$ given by
Eq.~\eqref{eq:Prob_coinc_double_MZI_gauss10} at the detectors
$D_6$ and $D_7$. The presence of multiple dips shows that the same
behavior can be obtained with or without ``overlapping photon wave
packets'' at the beam splitter $\text{BS}_2$. For this simulation
we used $\sigma\omega=7$ and $\tau_l\omega\in[-2\pi,2\pi]$ with
$l=1,2$.} \label{fig:MATLAB_prob_of_coinc_2MZI}
\end{figure}

The general form of the probability of coincident counts is rather
complicated, but if we restrict to $\tau_d=0$,
Eq.~\eqref{eq:Prob_coinc_double_MZI_01} simplifies to (see details
in Appendix \ref{sec:app:Pc_11_tau_d_zero})
%
%
%
\begin{eqnarray}
\label{eq:Prob_coinc_double_MZI}
P_c\left(t_0,\tau_1,\tau_2\right)=\frac{1}{16}\Big\vert\left[\zeta_0\left(t_0\right)+\zeta_0\left(t_0-\tau_1\right)\right]
\qquad
\nonumber\\ 
\cdot\left[\zeta_1\left(t_0\right)-\zeta_1\left(t_0-\tau_1\right)\right]
\nonumber\\ 
-\left[\zeta_0\left(t_0-\tau_2\right)-\zeta_0\left(t_0-\tau_1-\tau_2\right)\right]
\qquad
\nonumber\\ 
\cdot\left[\zeta_1\left(t_0-\tau_2\right)+\zeta_1\left(t_0-\tau_1-\tau_2\right)\right]\Big\vert^2
\end{eqnarray}
%
%
If we employ the spatio-temporal modes from
Eqs.~\eqref{eq:zeta0_Gaussian} and \eqref{eq:zeta1_Gaussian},
after a series of rather long calculations (detailed in Appendix
\ref{sec:app:Pc_11_gaussian_calculation}), the time-integrated
(over $t_0$ and $\tau_d$) probability  of coincidence for
$\tau_e=0$ yields
\begin{eqnarray}
\label{eq:Prob_coinc_double_MZI_gauss10}
P_c\left(\tau_1,\tau_2\right)=\frac{1}{8}\Big(2
-2\text{e}^{-\frac{\tau_1^2}{\sigma^2}}\cos\left(2\omega\tau_1\right)
\qquad\qquad\qquad\nonumber\\
-2\text{e}^{-\frac{\tau_2^2}{\sigma^2}}\cos\left(2\omega\tau_2\right)
+\text{e}^{-\frac{(\tau_1+\tau_2)^2}{\sigma^2}}\cos\left(2\omega(\tau_1+\tau_2)\right)
\nonumber\\
+\text{e}^{-\frac{(\tau_1-\tau_2)^2}{\sigma^2}}\cos\left(2\omega(\tau_1-\tau_2)\right)
\Big)
\end{eqnarray}
The surface plot of $P_c$ from
Eq.~(\ref{eq:Prob_coinc_double_MZI_gauss10}) versus $\tau_1$ and
$\tau_2$ is depicted in Fig. \ref{fig:MATLAB_prob_of_coinc_2MZI}.
As expected, for $\tau_1=\tau_2=0$, we have a dip in the
probability of coincident counts. But besides this dip, there are
other minima of $P_c$ for $\tau_1\neq0$ and $\tau_2\neq0$. This
time, we have no possible ``photon wave packets'' overlapping at
the beam splitter $\text{BS}_2$.

If we consider the transition from narrowband to monochromatic
light quanta (\emph{i.e.} when $\sigma\to\infty$),
Eq.~\eqref{eq:Prob_coinc_double_MZI_gauss10} becomes
$P_c\left(\tau_1,\tau_2\right)=\sin^2\left(\omega\tau_1\right)\sin^2\left(\omega\tau_2\right)$,
\emph{i.e.} we find the same result from
Eq.~\eqref{eq:double_MZI_Pc_monochromatic}.

\section{Conclusions}
\label{sec:conclusions} In this paper we introduced and analyzed a
new experimental setup within the standard formalism of QO. The
proposed double Mach-Zehnder interferometer experiment shows a
counter-intuitive behavior when excited with a single light
quantum: the path length difference of one MZI ``modulates'' the
amplitude of singles detection rates versus the path length
difference of the other MZI. In the extreme case, for certain
values of the path length difference of one MZI, the output
singles detection rates do not change, whatever the length of the
other MZI.

In the case of two simultaneously impinging light quanta at its
input, the double MZI experiment is able to show the same ``HOM
dip'' behavior for certain values of the delays $\varphi_1$ and
$\varphi_2$, whether or not the ``photon wave packets'' meet at
the second beam spitter. In the extreme case of nearly
monochromatic light, $\varphi_1$ and $\varphi_2$ may introduce
arbitrarily large delays, nonetheless the ``HOM dip'' in the
coincident counts will be present, showing that the intuitive
image of ``overlapping photon wave packets'' at the beam splitter
has a questionable physical reality, not being the key to explain
this experiment.

\appendix

\section{Computation of the input field operators in respect with the output ones}
\label{sec:app:compute_a0_a1_dagger}
We write down the ``inverse'' input-output operator relations for
each beam splitter. Starting with $\text{BS}_1$ we have
\begin{equation}
\label{eq:app:a1_dagger_fct_a2_a3}
\hat{a}_1^\dagger=\frac{1}{\sqrt{2}}\left(i\hat{a}_2^\dagger+\hat{a}_3^\dagger\right)
\end{equation}
and
\begin{equation}
\label{eq:app:a0_dagger_fct_a2_a3}
\hat{a}_0^\dagger=\frac{1}{\sqrt{2}}\left(\hat{a}_2^\dagger+i\hat{a}_3^\dagger\right)
\end{equation}
For the second beam splitter we also take into account the delay
$\varphi_1$, yielding
\begin{equation}
\label{eq:app:a2_dagger_fct_a4_a5}
\hat{a}_2^\dagger=\frac{1}{\sqrt{2}}\left(\hat{a}_4^\dagger+i\hat{a}_5^\dagger\right)
\end{equation}
and
\begin{equation}
\label{eq:app:a3_dagger_fct_a4_a5_delay1}
\hat{a}_3^\dagger=\text{e}^{-i\varphi_1}\frac{1}{\sqrt{2}}\left(i\hat{a}_4^\dagger+\hat{a}_5^\dagger\right)
\end{equation}
Finally, the beam splitter $\text{BS}_3$ transforms the field
operators
\begin{equation}
\label{eq:app:a5_dagger_fct_a6_a7}
\hat{a}_5^\dagger=\frac{1}{\sqrt{2}}\left(i\hat{a}_6^\dagger+\hat{a}_7^\dagger\right)
\end{equation}
and
\begin{equation}
\label{eq:app:a4_dagger_fct_a6_a7_delay1}
\hat{a}_4^\dagger=\text{e}^{-i\varphi_2}\frac{1}{\sqrt{2}}\left(\hat{a}_6^\dagger+i\hat{a}_7^\dagger\right)
\end{equation}
where in the last equation we took into account the delay
introduced by $\varphi_2$. Combining
Eqs.~\eqref{eq:app:a2_dagger_fct_a4_a5} and
\eqref{eq:app:a3_dagger_fct_a4_a5_delay1} into
Eq.~\eqref{eq:app:a0_dagger_fct_a2_a3}, we obtain
\begin{equation}
\hat{a}_0^\dagger
=\frac{1}{2}\left(\left(1-\text{e}^{-i\varphi_1}\right)\hat{a}_4^\dagger
+i\left(1+\text{e}^{-i\varphi_1}\right)\hat{a}_5^\dagger\right)
\end{equation}
and replacing $\hat{a}_4^\dagger$ and $\hat{a}_5^\dagger$ from
\eqref{eq:app:a5_dagger_fct_a6_a7} and, respectively,
\eqref{eq:app:a4_dagger_fct_a6_a7_delay1} yields
\begin{eqnarray}
\hat{a}_0^\dagger
=\frac{1}{2\sqrt{2}}\Big(\left(1-\text{e}^{-i\varphi_1}\right)
\text{e}^{-i\varphi_2}\left(\hat{a}_6^\dagger+i\hat{a}_7^\dagger\right)
\nonumber\\ 
+i\left(1+\text{e}^{-i\varphi_1}\right)\left(i\hat{a}_6^\dagger+\hat{a}_7^\dagger\right)\Big)
\end{eqnarray}
We group together the $\hat{a}_6^\dagger$ and $\hat{a}_7^\dagger$
terms, arriving at the final expression
\begin{eqnarray}
\label{eq:a0_dagger_fct_a6_a7_non_factored_doubleMZI}
\hat{a}_0^\dagger
=\frac{\text{e}^{-i\varphi_2}-\text{e}^{-i\varphi_1}\text{e}^{-i\varphi_2}-1-\text{e}^{-i\varphi_1}}{2\sqrt{2}}
\hat{a}_6^\dagger
\nonumber\\ 
+\frac{\text{e}^{-i\varphi_2}-\text{e}^{-i\varphi_1}\text{e}^{-i\varphi_2}+1+\text{e}^{-i\varphi_1}}{2\sqrt{2}}i\hat{a}_7^\dagger
\end{eqnarray}
Factoring out a common phase factor and replacing the complex
exponentials with trigonometric functions takes us to
Eq.~\eqref{eq:a0_dagger_in_respect_w_a6_a7}. Similar computations
lead us to
\begin{eqnarray}
\label{eq:a1_dagger_fct_a6_a7_non_factored_doubleMZI}
\hat{a}_1^\dagger
=\frac{\text{e}^{-i\varphi_2}+\text{e}^{-i\varphi_1}\text{e}^{-i\varphi_2}-1+\text{e}^{-i\varphi_1}}{2\sqrt{2}}i\hat{a}_6^\dagger
\nonumber\\ 
+\frac{-\text{e}^{-i\varphi_2}-\text{e}^{-i\varphi_1}\text{e}^{-i\varphi_2}-1+\text{e}^{-i\varphi_1}}{2\sqrt{2}}\hat{a}_7^\dagger
\end{eqnarray}
and by the same factorization we end up with
Eq.~\eqref{eq:a1_dagger_in_respect_w_a6_a7}. Please note that in
Eqs.~\eqref{eq:a0_dagger_in_respect_w_a6_a7} and
\eqref{eq:a1_dagger_in_respect_w_a6_a7} we chose to factor out
$\varphi_2$ (\emph{i.e.} we ended up with expressions involving
sines and cosines of $\varphi_2$). The same operations could have
been done with $\varphi_1$. It can be shown that none of the
observables (\emph{e.g.} singles, coincidence detection rates
etc.) changes.

\section{Computation of the output field operators $\hat{E}_6^{(+)}(t)$ and $\hat{E}_7^{(+)}(t)$ for the double MZI experiment}
\label{sec:app:compute_E6_E7_plus} We will deduce only the field
operator $\hat{E}_7^{(+)}\left(t\right)$ since the computation for
$\hat{E}_6^{(+)}\left(t\right)$ is following a similar path. We
start with the beam splitter $\text{BS}_3$ and the output field
operator $\hat{E}_7^{(+)}\left(t\right)$, connected to the input
field operators by
\begin{equation}
\label{eq:E_7_plus_fct_E4_E5}
\hat{E}_7^{(+)}\left(t\right)=\frac{1}{\sqrt{2}}\left(i\hat{E}_4^{(+)}\left(t-\tau_2\right)+\hat{E}_5^{(+)}\left(t\right)\right)
\end{equation}
where in the first term we took into account the delay
$\varphi_2$. We replace now the field operators for the beam
splitter $\text{BS}_2$, yielding
\begin{equation}
\hat{E}_4^{(+)}\left(t\right)=\frac{1}{\sqrt{2}}\left(\hat{E}_2^{(+)}\left(t\right)+i\hat{E}_3^{(+)}\left(t\right)\right)
\end{equation}
and
\begin{equation}
\hat{E}_5^{(+)}\left(t\right)=\frac{1}{\sqrt{2}}\left(i\hat{E}_2^{(+)}\left(t\right)+\hat{E}_3^{(+)}\left(t\right)\right)
\end{equation}
We can replace now these operators into
\eqref{eq:E_7_plus_fct_E4_E5} yielding for our field operator
\begin{eqnarray}
\label{eq:E_7_plus_fct_E2_E2}
\hat{E}_7^{(+)}\left(t\right)=\frac{1}{2}\Big(i\hat{E}_2^{(+)}\left(t-\tau_2\right)
-\hat{E}_3^{(+)}\left(t-\tau_1-\tau_2\right)
\nonumber\\
+i\hat{E}_2^{(+)}\left(t\right)+\hat{E}_3^{(+)}\left(t-\tau_1\right)\Big)
\end{eqnarray}
where we took into account the delay $\varphi_1$ for the field
operator $\hat{E}_3^{(+)}\left(t\right)$. Combining
Eqs.~\eqref{eq:E_2_plus_fct_zet_0_zeta_1} and
\eqref{eq:E_3_plus_fct_zet_0_zeta_1} with
Eq.~\eqref{eq:E_7_plus_fct_E2_E2}, by simply regrouping some terms
we arrive at the desired result
Eq.~\eqref{eq:E7__non_monochrom_tau_e_tau_d}.

\section{Computation of the probability of singles detection for the double MZI experiment}
\label{sec:app:P6_single_photon_detection_2MZI}

For an input state
$\vert\psi_{in}\rangle=\vert0_01_1\rangle=\hat{a}_1^\dagger\vert0\rangle$,
the probability of photo-detection at the detector $D_6$ can be
written as
\begin{equation}
\label{eq:app:P6_double_MZI_single}
P_6=\Vert\hat{E}^+_6\left(t_0\right)\hat{a}_1^\dagger\vert0\rangle\Vert^2
\end{equation}
This expression can be simplified if we note that only terms
containing $\hat{a}_1$ will yield a contribution to $P_6$ since
$\hat{a}_0\hat{a}_1^\dagger\vert0\rangle=0$. Inserting
Eq.~\eqref{eq:E6__non_monochrom_tau_e_tau_d} into
Eq.~\eqref{eq:app:P6_double_MZI_single} leads to
\begin{equation}
\label{eq:app:P6_double_MZI_single_bis}
P_6=\frac{1}{8}\Big\vert\zeta_1\left(t-\tau_2\right)+\zeta_1\left(t-\tau_1-\tau_2\right)
-\zeta_1\left(t\right)+\zeta_1\left(t-\tau_1\right)\Big\vert^2
\end{equation}
We can replace now $\zeta_1\left(t\right)$ with the Gaussian mode
defined in Eq.~\eqref{eq:zeta1_Gaussian}. Since $\tau_e$ has no
meaning for a single light quantum, we set it to zero. Expanding
the mod-square from Eq.~\eqref{eq:app:P6_double_MZI_single_bis}
yields

\begin{widetext}

\begin{eqnarray}
P_6=\frac{1}{8}\sqrt{\frac{2}{\pi\sigma^2}}\bigg(
\text{e}^{-2\frac{(t_0-\tau_2)^2}{\sigma^2}}
+\text{e}^{-\frac{(t_0-\tau_2)^2}{\sigma^2}}\text{e}^{-\frac{(t_0-\tau_1-\tau_2)^2}{\sigma^2}}\text{e}^{-i\omega\tau_1}
-\text{e}^{-\frac{(t_0-\tau_2)^2}{\sigma^2}}\text{e}^{-\frac{t_0^2}{\sigma^2}}\text{e}^{i\omega\tau_2}
+\text{e}^{-\frac{(t_0-\tau_2)^2}{\sigma^2}}\text{e}^{-\frac{(t_0-\tau_1)^2}{\sigma^2}}\text{e}^{i\omega(\tau_2-\tau_1)}
\nonumber\\ 
+\text{e}^{-\frac{(t_0-\tau_1-\tau_2)^2}{\sigma^2}}\text{e}^{-\frac{(t_0-\tau_2)^2}{\sigma^2}}\text{e}^{i\omega\tau_1}
+\text{e}^{-2\frac{(t_0-\tau_1-\tau_2)^2}{\sigma^2}}
-\text{e}^{-\frac{(t_0-\tau_1-\tau_2)^2}{\sigma^2}}\text{e}^{-\frac{t_0^2}{\sigma^2}}\text{e}^{i\omega(\tau_1+\tau_2)}
+\text{e}^{-\frac{(t_0-\tau_1-\tau_2)^2}{\sigma^2}}\text{e}^{-\frac{(t_0-\tau_1)^2}{\sigma^2}}\text{e}^{i\omega\tau_2}
\nonumber\\
-\text{e}^{-\frac{t_0^2}{\sigma^2}}\text{e}^{-\frac{(t_0-\tau_2)^2}{\sigma^2}}\text{e}^{-i\omega\tau_2}
-\text{e}^{-\frac{t_0^2}{\sigma^2}}\text{e}^{-\frac{(t_0-\tau_1-\tau_2)^2}{\sigma^2}}\text{e}^{-i\omega(\tau_1+\tau_2)}
+\text{e}^{-2\frac{t_0^2}{\sigma^2}}-\text{e}^{-\frac{t_0^2}{\sigma^2}}\text{e}^{-\frac{(t_0-\tau_1)^2}{\sigma^2}}\text{e}^{-i\omega\tau_1}
\nonumber\\
+\text{e}^{-\frac{(t_0-\tau_1)^2}{\sigma^2}}\text{e}^{-\frac{(t_0-\tau_2)^2}{\sigma^2}}\text{e}^{i\omega(\tau_1-\tau_2)}
+\text{e}^{-\frac{(t_0-\tau_1)^2}{\sigma^2}}\text{e}^{-\frac{(t_0-\tau_1-\tau_2)^2}{\sigma^2}}\text{e}^{-i\omega\tau_2}
-\text{e}^{-\frac{(t_0-\tau_1)^2}{\sigma^2}}\text{e}^{-\frac{t_0^2}{\sigma^2}}\text{e}^{i\omega\tau_1}
+\text{e}^{-2\frac{(t_0-\tau_1)^2}{\sigma^2}}\bigg)\quad
\end{eqnarray}
In order to $t_0$-integrate the expression above we use the
formula
\begin{equation}
\label{eq:integral_tsquare_minus_a_minus_b_sigma_over_2}
\int_{-\infty}^{\infty}{\text{e}^{-\frac{(y-a)^2}{\sigma^2}}\text{e}^{-\frac{(y-b)^2}{\sigma^2}}\text{d}y}
=\sqrt{\frac{\pi\sigma^2}{2}}\text{e}^{-\frac{(a-b)^2}{2\sigma^2}}
\end{equation}
valid for any $a,b\in\mathbb{R}$ and $\sigma\in\mathbb{R}^*$.
After time-integration and some simplifications we are left with
\begin{eqnarray}
P_6=\frac{1}{8}\left(
4+\text{e}^{-\frac{(\tau_1-\tau_2)^2}{2\sigma^2}}\text{e}^{i\omega(\tau_2-\tau_1)}
+\text{e}^{-\frac{(\tau_1-\tau_2)^2}{2\sigma^2}}\text{e}^{-i\omega(\tau_2-\tau_1)}
-\text{e}^{-\frac{(\tau_1+\tau_2)^2}{2\sigma^2}}\text{e}^{i\omega(\tau_1+\tau_2)}
-\text{e}^{-\frac{(\tau_1+\tau_2)^2}{2\sigma^2}}\text{e}^{-i\omega(\tau_1+\tau_2)}
 \right)\qquad
\end{eqnarray}
Grouping the complex exponentials into cosines takes us to the
final result from
Eq.~\eqref{eq:P6_2MZI_non_monochrom_10_input_state}. The
computation of $P_7$ follows identical steps.

\section{Computation of the probability of coincident counts for the double MZI experiment}
\label{sec:app:Pc_11_tau_d_zero} The probability of coincident
counts from Eq.~\eqref{eq:Prob_coinc_double_MZI_01} can be written
as
\begin{equation}
\label{eq:Prob_coinc_double_MZI_general_gauss_start_app}
P_c\left(t_0,\tau_1,\tau_2,\tau_d\right)=
\Vert\hat{E}_7^{(+)}\left(t_0+\tau_d\right)\hat{E}_6^{(+)}\left(t_0\right)\hat{a}_1^\dagger\hat{a}_0^\dagger\vert0\rangle\Vert^2
\end{equation}
and we can readily discard from
Eq.~\eqref{eq:Prob_coinc_double_MZI_general_gauss_start_app}
all terms containing $\hat{a}_1^2$ and $\hat{a}_0^2$ since
$\hat{a}_1^2\hat{a}_1^\dagger\hat{a}_0^\dagger\vert0\rangle=\hat{a}_0^2\hat{a}_1^\dagger\hat{a}_0^\dagger\vert0\rangle=0$.
We are left with the rather complicated expression
\begin{eqnarray}
\label{eq:Prob_coinc_double_MZI_general_gauss_3}
P_c\left(t_0,\tau_1,\tau_2,\tau_d\right)
=\frac{1}{64}\bigg\vert
\Big(\zeta_0\left(t_0-\tau_2\right)-\zeta_0\left(t_0-\tau_1-\tau_2\right)
-\zeta_0\left(t_0\right)-\zeta_0\left(t_0-\tau_1\right)\Big)\nonumber\\
\cdot\Big(-\zeta_1\left(t_0-\tau_2+\tau_d\right)-\zeta_1\left(t_0-\tau_1-\tau_2+\tau_d\right)
-\zeta_1\left(t_0+\tau_d\right)+\zeta_1\left(t_0-\tau_1+\tau_d\right)\Big)\nonumber\\
-\Big(\zeta_0\left(t_0-\tau_2+\tau_d\right)-\zeta_0\left(t_0-\tau_1-\tau_2+\tau_d\right)
+\zeta_0\left(t_0+\tau_d\right)+\zeta_0\left(t_0-\tau_1+\tau_d\right)\Big)\nonumber\\
\cdot\Big(\zeta_1\left(t_0-\tau_2\right)+\zeta_1\left(t_0-\tau_1-\tau_2\right)
-\zeta_1\left(t\right)+\zeta_1\left(t_0-\tau_1\right)\Big)
\bigg\vert^2
\end{eqnarray}
However, for $\tau_d=0$ half of the terms disappear while the
other half becomes pairs of identical terms yielding
\begin{eqnarray}
\label{eq:Prob_coinc_double_MZI_general_gauss_7}
P_c\left(t_0,\tau_1,\tau_2\right)=\frac{1}{16}\Big\vert
-\zeta_0\left(t_0-\tau_2\right)\zeta_1\left(t_0-\tau_2\right)
-\zeta_0\left(t_0-\tau_2\right)\zeta_1\left(t_0-\tau_1-\tau_2\right)
\qquad\qquad
\nonumber\\
+\zeta_0\left(t_0-\tau_1-\tau_2\right)\zeta_1\left(t_0-\tau_2\right)
+\zeta_0\left(t_0-\tau_1-\tau_2\right)\zeta_1\left(t_0-\tau_1-\tau_2\right)
+\zeta_0\left(t_0\right)\zeta_1\left(t_0\right)
\nonumber\\
-\zeta_0\left(t_0\right)\zeta_1\left(t_0-\tau_1\right)
+\zeta_0\left(t_0-\tau_1\right)\zeta_1\left(t_0\right)
-\zeta_0\left(t_0-\tau_1\right)\zeta_1\left(t_0-\tau_1\right)
\Big\vert^2
\end{eqnarray}
and after some basic algebra, this expression dramatically
simplifies yielding Eq.~\eqref{eq:Prob_coinc_double_MZI}.

\section{Computation of the time-integrated probability of coincident counts for non-monochromatic light quanta}
\label{sec:app:Pc_11_gaussian_calculation} We replace into
Eq.~\eqref{eq:Prob_coinc_double_MZI} the Gaussian waveforms given
by Eqs.~\eqref{eq:zeta0_Gaussian} and \eqref{eq:zeta1_Gaussian}
and set $\tau_e=0$ \emph{i.e.} we have simultaneously impinging
light quanta on beam splitter $\text{BS}_1$. Expanding now the
mod-square yields
\begin{eqnarray}
\label{eq:Prob_coinc_double_MZI_gauss8_sigma}
P_c\left(t_0,\tau_1,\tau_2\right)=\frac{1}{16}\frac{2}{\pi\sigma^2}\Big(
\text{e}^{-4\frac{t_0^2}{\sigma^2}}
-\text{e}^{-2\frac{t_0^2}{\sigma^2}}\text{e}^{-2\frac{(t_0-\tau_1)^2}{\sigma^2}}\text{e}^{-2i\omega\tau_1}
-\text{e}^{-2\frac{t_0^2}{\sigma^2}}\text{e}^{-2\frac{(t_0-\tau_2)^2}{\sigma^2}}\text{e}^{-2i\omega\tau_2}
\qquad\qquad\qquad\qquad
\nonumber\\
+\text{e}^{-2\frac{t_0^2}{\sigma^2}}\text{e}^{-2\frac{(t_0-\tau_1-\tau_2)^2}{\sigma^2}}\text{e}^{-2i\omega(\tau_1+\tau_2)}
-\text{e}^{-2\frac{t_0^2}{\sigma^2}}\text{e}^{-2\frac{(t_0-\tau_1)^2}{\sigma^2}}\text{e}^{2i\omega\tau_1}
+\text{e}^{-4\frac{(t_0-\tau_1)^2}{\sigma^2}}
+\text{e}^{-2\frac{(t_0-\tau_2)^2}{\sigma^2}}\text{e}^{-2\frac{(t_0-\tau_1)^2}{\sigma^2}}\text{e}^{2i\omega(\tau_1-\tau_2)}
\nonumber\\
-\text{e}^{-2\frac{(t_0-\tau_1)^2}{\sigma^2}}\text{e}^{-2\frac{(t_0-\tau_1-\tau_2)^2}{\sigma^2}}\text{e}^{-2i\omega\tau_2}
-\text{e}^{-2\frac{t_0^2}{\sigma^2}}\text{e}^{-2\frac{(t_0-\tau_2)^2}{\sigma^2}}\text{e}^{2i\omega\tau_2}
+\text{e}^{-2\frac{(t_0-\tau_2)^2}{\sigma^2}}\text{e}^{-2\frac{(t_0-\tau_1)^2}{\sigma^2}}\text{e}^{2i\omega(\tau_2-\tau_1)}
+\text{e}^{-4\frac{(t_0-\tau_2)^2}{\sigma^2}}
\nonumber\\
-\text{e}^{-2\frac{(t_0-\tau_2)^2}{\sigma^2}}\text{e}^{-2\frac{(t_0-\tau_1-\tau_2)^2}{\sigma^2}}\text{e}^{-2i\omega\tau_1}
+\text{e}^{-2\frac{t_0^2}{\sigma^2}}\text{e}^{-2\frac{(t_0-\tau_1-\tau_2)^2}{\sigma^2}}\text{e}^{2i\omega(\tau_1+\tau_2)}
-\text{e}^{-2\frac{(t_0-\tau_1)^2}{\sigma^2}}\text{e}^{-2(t_0-\tau_1-\tau_2)^2}\text{e}^{2i\omega\tau_2}
\nonumber\\
-\text{e}^{-2\frac{(t_0-\tau_2)^2}{\sigma^2}}\text{e}^{-2(t_0-\tau_1-\tau_2)^2}\text{e}^{2i\omega\tau_1}
+\text{e}^{-4\frac{(t_0-\tau_1-\tau_2)^2}{\sigma^2}} \Big)\quad
\end{eqnarray}
In order to perform the time integration of this expression we
will be using the formula
\begin{equation}
\label{eq:integral_tsquare_minus_a_minus_b_sigma}
\int_{-\infty}^{\infty}{\text{e}^{-2\frac{(y-a)^2}{\sigma^2}}\text{e}^{-2\frac{(y-b)^2}{\sigma^2}}\text{d}y}
=\frac{\sigma\sqrt{\pi}}{2}\text{e}^{-\frac{(a-b)^2}{\sigma^2}}
\end{equation}
valid for any $a,b\in\mathbb{R}$ and $\sigma\in\mathbb{R}^*$.
Time-integrating Eq.~\eqref{eq:Prob_coinc_double_MZI_gauss8_sigma}
and using cosines instead of complex exponentials takes us to
\begin{eqnarray}
\label{eq:Prob_coinc_double_MZI_gauss10_sigma}
P_c\left(\tau_1,\tau_2\right)=\frac{1}{8}\frac{1}{\sigma\sqrt{\pi}}\Big(2
-2\text{e}^{-\frac{\tau_1^2}{\sigma^2}}\cos\left(2\omega\tau_1\right)
-2\text{e}^{-\frac{\tau_2^2}{\sigma^2}}\cos\left(2\omega\tau_2\right)
\qquad\qquad\qquad\qquad\qquad\qquad
\nonumber\\
\qquad\qquad\qquad\qquad\qquad\qquad
+\text{e}^{-\frac{(\tau_1+\tau_2)^2}{\sigma^2}}\cos\left(2\omega(\tau_1+\tau_2)\right)
+\text{e}^{-\frac{(\tau_1-\tau_2)^2}{\sigma^2}}\cos\left(2\omega(\tau_1-\tau_2)\right)
\Big)
\end{eqnarray}
In order to obtain the $\tau_d$-integrated expression we should
have tackled Eq.~\eqref{eq:Prob_coinc_double_MZI_general_gauss_3},
an expression far too complicated for manual computation.
Nonetheless, we can reasonably assume that each term from
Eq.~\eqref{eq:Prob_coinc_double_MZI_gauss10_sigma} has a
$\text{e}^{-\tau_d^2/\sigma^2}$ factor that was ignored (since we
considered only the case $\tau_d=0$), therefore, if we were to
perform the time-integration over $\tau_d$ in the general case, we
should get an extra factor of $\sigma\sqrt{\pi}$, yielding the
probability of coincident counts
\begin{eqnarray}
\label{eq:Prob_coinc_double_MZI_gauss10_app}
P_c\left(\tau_1,\tau_2\right)=\frac{1}{8}\Big(2
-2\text{e}^{-\frac{\tau_1^2}{\sigma^2}}\cos\left(2\omega\tau_1\right)
-2\text{e}^{-\frac{\tau_2^2}{\sigma^2}}\cos\left(2\omega\tau_2\right)
\qquad\qquad\qquad\qquad\qquad\qquad
\nonumber\\
\qquad\qquad\qquad\qquad\qquad\qquad
+\text{e}^{-\frac{(\tau_1+\tau_2)^2}{\sigma^2}}\cos\left(2\omega(\tau_1+\tau_2)\right)
+\text{e}^{-\frac{(\tau_1-\tau_2)^2}{\sigma^2}}\cos\left(2\omega(\tau_1-\tau_2)\right)
\Big)
\end{eqnarray}
At this final stage, this expression was obtained by guess and
plausibility arguments, nonetheless the full expression of
Eq.~\eqref{eq:Prob_coinc_double_MZI_general_gauss_3} was
implemented on a computing machine, numerically time-integrated
(over $t_0$ and $\tau_d$) then compared to
Eq.~\eqref{eq:Prob_coinc_double_MZI_gauss10_app}. No differences
were found.

\end{widetext}


\end{document}